# Principles alone cannot guarantee ethical AI

Brent Mittelstadt[i]

**Abstract**

AI Ethics is now a global topic of discussion in academic and policy circles. At least 84 public-private initiatives have produced statements describing high-level principles, values, and other tenets to guide the ethical development, deployment, and governance of AI. According to recent meta-analyses, AI Ethics has seemingly converged on a set of principles that closely resemble the four classic principles of medical ethics. Despite the initial credibility granted to a principled approach to AI Ethics by the connection to principles in medical ethics, there are reasons to be concerned about its future impact on AI development and governance. Significant differences exist between medicine and AI development that suggest a principled approach in the latter may not enjoy success comparable to the former. Compared to medicine, AI development lacks (1) common aims and fiduciary duties, (2) professional history and norms, (3) proven methods to translate principles into practice, and (4) robust legal and professional accountability mechanisms. These differences suggest we should not yet celebrate consensus around high-level principles that hide deep political and normative disagreement.

## 1  Introduction

Over the past several years a plethora of public-private initiatives have arisen globally to define values, principles, and frameworks for the ethical development and deployment of AI.[1] These initiatives can help focus public debate on a common set of issues and principles, and raise awareness among the public, developers and institutions of the ethical challenges that accompany AI.[2]

To date, at least 84 such 'AI Ethics' initiatives have published reports describing high-level ethical principles, tenets, values, or other abstract requirements for AI development and deployment.[3] Many envision these high-level contributions being 'translated' into mid- or low-level design requirements and technical fixes, governance frameworks, and professional codes.[1]

Existing initiatives to codify AI Ethics are not without their critics. Many initiatives, particularly those sponsored by industry, have been characterised as mere virtue signalling intended to delay regulation and pre-emptively focus debate on abstract problems and technical solutions.[1,4] This view is difficult to dismiss: AI Ethics initiatives have thus far largely produced vague, high-level principles and value statements which promise to be action-guiding, but in practice provide few specific recommendations[5] and fail to address fundamental normative and political tensions embedded in key concepts (e.g. fairness, privacy). Declarations by AI companies and developers committing themselves to high-level ethical principles and self-regulatory codes nonetheless provide policy-makers with a reason not to pursue new regulation.[5,6]

Comparisons have recently been drawn between AI Ethics initiatives and medical ethics.[7] A recent review found that many AI Ethics initiatives have converged on a set of principles that closely resemble the four classic principles of medical ethics.[8] This finding has been subsequently endorsed by the OECD[9] and the European Commission's High Level Expert Group on Artificial Intelligence (HLEG), which proposed four principles to guide the development of 'trustworthy' AI: respect for human autonomy, prevention of harm, fairness, and explicability.[10]

[i] Oxford Internet Institute, University of Oxford, 1 St. Giles, Oxford, OX1 3JS, UK; the Alan Turing Institute, British Library, 96 Euston Road, London, NW1 2DB, UK. Correspondence via brent.mittelstadt@oii.ox.ac.uk.

This convergence of AI Ethics around principles of medical ethics is opportune, as it is historically the most prominent and well-studied approach to applied ethics. 'Principlism' emerged from medicine as a theoretical moral framework joining traditional ethical standards with the requirements of practitioners, research ethics committees, and medical institutions for practical ethical decision-making.[11] Principlism proposes four core principles that require specification and balancing in different decision-making contexts.[12] Whereas principlism in medical ethics provides a common language to identify and conceptualise ethical challenges,[13,14] and provides guidance for setting health policy and clinical decision-making, a principled approach in AI Ethics seems intended to embed normative considerations in technology design and governance. Both approaches address how to embed principles in professional practice. Principlism thus provides a helpful backdrop to assess the potential for AI Ethics to enact real change in the development and deployment of AI.

Despite the initial credibility lent by the comparison with medical ethics, there are reasons to be concerned about the future impact of AI Ethics. Important differences exist between medicine (and other traditional professions[11]) and AI development that suggest a principled approach in the latter may not enjoy success comparable to the former.

This paper critically assesses the strategies and recommendations proposed by current AI Ethics initiatives. Outputs of existing AI Ethics initiatives were reviewed to determine their proposed strategy for embedding ethics into the development and governance of AI.[3] Prior work on the implementation and impact of principlism in medicine is used to critically assess the potential impact of a principled approach to AI ethics.

## 2 The challenges of a principled approach to AI Ethics

Four characteristics of AI development suggest a principled approach may have limited impact on design and governance. Compared to medicine, AI development lacks (1) common aims and fiduciary duties, (2) professional history and norms, (3) proven methods to translate principles into practice, and (4) robust legal and professional accountability mechanisms.

### 2.1 Common aims and fiduciary duties

Medicine is broadly guided by a common aim: to promote the health and well-being of the patient.[15] It is a defining quality of a profession for its practitioners to be part of a 'moral community' with common aims, values, and training.[16–18] The pursuit of a common goal facilitates a principled approach to ethical decision-making.[11] While there is much disagreement over the meaning of 'health', and how best to promote it in practice, the interests of patients and medical practitioners remain aligned at some fundamental level which encourages solidarity and trust.[19] By some accounts, practitioners have a moral obligation to advocate for their patient's interests against institutional interests.[18–20]

Comparable solidarity cannot be taken for granted in AI development. AI is largely developed by the private sector for deployment in public (e.g. criminal sentencing) and private (e.g. insurance) contexts. The fundamental aims of developers, users, and affected parties do not necessarily align. Developers often "work in an environment which constantly pressures them to cut costs, increase profit and deliver higher quality" systems, and face pressure from management to make decisions that prioritise company interests.[21,22] While health professionals undoubtedly face similar organisational pressures, they are not equivalent in degree. Unlike medicine, AI development does not serve the equivalent of a 'patient' whose interests are granted initial primacy in ethical decision-making. This lack of a common goal transforms ethical decision-making from a cooperative to a competitive process, which makes finding a balance between public and private interests more difficult in practice.

The implicit solidarity of medicine is formally recognised in professional codes of practice and regulatory frameworks that establish fiduciary duties towards patients.[20] Formal professions are distinguished from other vocations by fiduciary duties derived from the client-practitioner relationship,[23] which are mediated by common goals and values within the profession, and enforced through sanctions and self-governance (see: Table 1).[16,24,25] These characteristics facilitate a principled approach to ethical decision-making by requiring practitioners to promote their clients' best interests.

AI development is not a formal profession. Equivalent fiduciary relationships and complementary governance mechanisms do not exist for private sector AI developers.[5] AI developers do not commit to 'public service', which in other professions requires practitioners to uphold public interests in the face of competing business or managerial interests.[24] For AI or software deployed in the public sector, such a commitment may be implicit in institutional or political structures. The same cannot be said for the private sector. Companies and their employees have principal fiduciary duties towards their shareholders. Public interests are not granted primacy over commercial interests.

It could be argued that many medical practitioners and institutions face similar pressures. Hospitals, for example, must ensure their model of healthcare delivery is sustainable and balances the interests of individual patients against public health goals. Likewise, many new therapies are developed in the private sector.

These apparent similarities are, however, superficial. Public and private medical institutions are required to operate within strict regulatory frameworks that ensure the health and well-being of patients and research participants are not subsumed in pursuit of sustainability or profit.[13] AI institutions are of course also subject to regulation in certain sectors; data protection and privacy law, for example, constrain processing of the personal data necessary to train models. However, the impact of such frameworks is variable and limited in scope.[26,27] A unified regulatory framework does not yet exist for AI which establishes clear fiduciary duties towards data subjects and users. Should such a framework emerge from AI Ethics, a principled approach could be deemed successful. But in the absence of strong regulation that establishes fiduciary duties or primacy for the vital interests of data subjects and users,[28] a comparable degree of value alignment cannot be said to exist for AI.

The absence of a fiduciary relationship in AI means that users cannot trust that developers will act in their best interests when implementing ethical principles in practice. Reputational risks may push companies to engage with ethics, but these risks carry weight only as long as they remain in the public consciousness. Personal moral conviction may also push AI developers towards 'good' behaviour;[29] recent examples of internal protests at Google provide some cause for hope.[30,31] However, incentive structures that discourage 'whistleblowing' or placing public interests before the company suggest that virtuous actions will often come at a high personal cost.[32] It is unacceptable that users and affected parties must rely on the personal convictions of developers, fear of reputational damage, or public outcry for their vital interests in privacy, autonomy, identity, and other areas to be taken seriously.[33]

## 2.2 Professional history and norms

The second weakness of a principled approach to AI Ethics is a relative lack of professional history and well-defined norms of 'good' behaviour. Medicine benefits from a long history and shared professional culture with variation across cultures and specialties. Accounts of the moral obligations and virtues of health professions have developed over centuries.[14] In Western biomedicine, longstanding standards set in the Hippocratic Oath, the Declaration of Geneva, the Declaration of Helsinki, and other accounts of the 'good' health professional have served as a basis for clinical decision-making and research ethics,[34] and inspired the development of ethics and codes in other professions.[11]

These standards and accounts of being a 'good' doctor have not entirely prevented ethical negligence in clinical practice and medical research. Nonetheless, they provide a historically sensitive account of the obligations of the profession against which negligent conduct and practices can be identified.[19] As such failures have occurred across the profession's history, and as new technologies, treatments, and changing social values have disrupted established norms of good behaviour, ethical standards have been revised.

Evidence of these historical lessons can be seen in modern codes of conduct and ethics; the American Medical Association's Code of Medical Ethics, for example, is a highly detailed document detailing opinions, behavioural norms, and standards across a plethora of medical practices and technologies.[35] The first version (1847) of the code focused solely on professional conduct and emphasised a paternalistic duty to maximize benefits and minimize harms to patients. Over time the profession's standards have shifted away from this myopic focus on beneficence and professional conduct, and towards other duties owed to patients, most notably a growing respect for autonomy.[14,34] Particularly egregious failures have often provided the catalyst for change; atrocities in human experimentation, involuntary sterilization, and euthanasia committed in Nazi Germany during World War II, for example, led to the formalisation of a set of research ethics principled in the 1947 Nuremberg Code. Similarly, in the United States the National Research Act (1974) and Belmont Report (1978) were drafted in response to abuses observed in the Tuskegee Syphilis Study and other human studies.[13,14,34]

Principlism[12] subsequently arose to correct the "minimal and unsatisfactory professional morality" the Hippocratic tradition was perceived to have created.[14] To move beyond this tradition, "explicit recognition of basic ethical principles that could help identify various clinical practices and human experiments as morally questionable or unacceptable" was required.[14] By addressing concerns with conduct as well as practice, principlism has established a common moral language to identify and address problematic behaviours and practices. This framework has subsequently taken hold in curricula and training for medical students, practitioners, and policy-makers, and continues to influence ethical codes and decision-making in medicine.[13,14,36] Principlism thus has a strong, historically informed regulating influence on the behaviour and ethics of medical practitioners and institutions.

AI development does not have a comparable history, homogenous professional culture and identity, or similarly developed professional ethics frameworks. The profession has not gone through comparable transformative moments in which its ethical obligations are clearly recognised and translated into specific, practical moral duties and best practices. Whereas AI can in principle be deployed in any context involving human expertise, medicine in comparison has narrower aims which facilitates development of standard practices and norms. Reflecting this, AI developers come from varied disciplines and professional backgrounds which have incongruous histories, cultures, incentive structures, and moral obligations.[11,24,25] Reducing the field to a single vocation or type of expertise would be an oversimplification.[11]

Software engineering, which is arguably the closest analogue, has historically not been legally recognised as a profession with fiduciary duties to the public[11,37] due to the absence of licensure schemes and a well-defined professional 'standard of care'.[38] Reflecting this, a comparably rich account of what it means to be a 'good' AI developer or software engineer does not exist; while the IEEE and ACM, two of the field's largest professional associations, have published and revised codes of ethics, these documents remain comparatively short, theoretical, and lacking in grounded advice and specific behavioural norms.[11]

Stronger professional standards (and supporting organisations) for AI development could, of course, be drafted in the future. Unfortunately, this will not be a simple task. Systems are often created by large, multi-disciplinary and multi-national teams. Whereas the effects of clinical decision-making are often (but not always) immediate and observable, the impact of decisions taken in designing, training, and configuring AI systems for different uses may never become apparent to developers.[39] This is worrying, as distance from potential victims has been shown to have a positive effect on unethical professional behaviour.[40] The risks addressed by medical ethics largely arise from interventions performed (or not) on the physical body. In comparison, ethical risks in AI are continuous and not similarly bound, and may not be directly experienced by data subjects.[41] Systems are also often 'opaque' in the sense that no single person will have a full understanding of the system's design or functionality,[42] or be able to predict its behaviour. Even where problems are recognised, they can rarely be traced back to a single team member or action;[43] responsibility must be assigned across a network of actors that influenced the system's design, training, and configuration. This inability to reliably predict the effects of development choices undermines the creation of standards to be a 'good' AI developer or requirements for 'good' AI.

AI Ethics initiatives address this gap by defining broadly acceptable principles to guide the people and processes responsible for the development, deployment, and governance of AI across radically different contexts of use. At this level of abstraction meaningful guidance may be impossible.[44,45] The great diversity of stakeholders and interests involved necessarily pushes the search for common values and norms towards a high level of abstraction.[24] The results are statements of principles or values based on abstract and vague concepts, for example commitments to ensure AI is 'fair' or respects 'human dignity', which are not specific enough to be action-guiding.[7]

Statements reliant on vague normative concepts hide points of political and ethical conflict. 'Fairness', 'dignity', and other such abstract concepts are examples of "essentially contested concepts" which have many possible conflicting meanings requiring contextual interpretation through one's background political and philosophical beliefs.[46] These different interpretations, which can be rationally and genuinely held, lead to substantively different requirements in practice[47] which will only be revealed once principles or concepts are translated and tested in practice.[48] At best, this conceptual ambiguity allows for context-sensitive specification of ethical requirements for AI. At worst, it masks fundamental, principled disagreement and drives AI Ethics towards moral relativism. At a minimum, any compromise reached thus far around core principles for AI Ethics does not reflect meaningful consensus on a common practical direction for 'good' AI development and governance. We must not confuse high-level compromise with *a priori* consensus.

The truly difficult part of ethics—actually translating normative theories, concepts and values into 'good' practices AI practitioners can adopt—is kicked down the road like the proverbial can. Developers are left to translate principles and specify essentially contested concepts as they see fit, without a clear roadmap for unified implementation. This process will likely encounter incommensurable moral norms and frameworks which present true moral dilemmas that principlism cannot resolve.[49–51] An established profession that can draw on a rich history, ethical culture, and norms of 'good' practice will be best placed to conceptualise and debate (if not resolve) these challenges.[11,52] Unfortunately, AI development is left wanting in this regard.

## 2.3 Methods to translate principles in practice

The third weakness of a principled approach to AI Ethics is the absence of proven methods to translate principles into practice. The prevalence of essentially contested concepts in AI Ethics begs a question: How can normative disagreements over the 'correct' specification of such concepts be resolved?

Principles do not automatically translate into practice.[21] Throughout its history, medicine has developed effective ways of translating high-level commitments and principles into practical requirements and norms of 'good' practice.[14,18] Professional societies and boards, ethics review committees, accreditation and licensing schemes, peer self-governance, codes of conduct and other mechanisms supported by strong institutions help determine the ethical acceptability of day-to-day practice by assessing difficult cases, identifying negligent behaviour, and sanctioning bad actors.[53,54] The formal codes and informal norms that govern medical practice have been extensively tested, studied, and revised over time, with their recommendations and norms (and underlying principles) evolving to remain relevant. High-level principles rarely feature explicitly in clinical decision-making which is instead bounded by institutional policies that include principled concerns.[13] On the ground, case-relevant precedents or specifications which use the moral language provided by principlism are more common.[36,55] Taken together, institutional and clinical decision-making are effectively a 'coherence' approach accounting for both high-level and grounded considerations.[12]

AI development does not have comparable empirically proven methods to translate principles into practice in real-world development contexts. This is a multi-faceted methodological challenge. Translation involves the specification of high-level principles into mid-level norms and low-level requirements. Norms and requirements cannot be deduced directly from mid-level principles without accounting for specific elements of the technology, application, context of use, or relevant local norms.[51,55,56] Normative decisions must be made at each stage of translation, and coherence must be sought between principles, specified norms or rules, and the facts of the case.[12,56] It follows that the justification arising from widespread consensus on a set of common principles does not transfer to the mid-level norms and low-level requirements derived from them. Each stage of translation and specification must be independently justified. A universally accepted hierarchy of principles to prioritise competing norms does not exist.[49,57]

This observation reveals the scope of work that remains for AI Ethics. High-level consensus is encouraging, but it has little bearing on the justification of norms and practical requirements proposed within specific contexts of use. Due to the necessity of local specification and justification, the prominence of essentially contested concepts in AI Ethics, and the field's relative lack of a binding professional history (see: Section 2.2), conflicting practical requirements will almost certainly emerge across the diverse sectors and contexts in which a principled approach to AI Ethics is used.

One other methodological challenge remains. Normative practical requirements must somehow be embedded in development processes, and functionally implanted in design requirements. Prior work points to the difficulty of embedding ethical values and principles in technology design and the development cycle.[1,56,58,59] Many such methods exist,[60] including participatory design, reflective design, Values@Play, and Value-Sensitive Design,[1] but thus far they have largely been implemented and studied in academic contexts which are more receptive to normative concerns than commercial settings.[22,60,61] Value-conscious methods are also largely procedural, not functional. Generally speaking, they introduce values, normative issues, and relevant stakeholders into the development process.[56] They do not, however, allow for particular values to be 'injected' into system design, and struggle to capture the degree to which the resulting artefact reflects particular values or specifications.[58]

Value-conscious design frameworks face additional challenges in commercial development processes. Ethics has a cost. AI is often developed behind 'closed doors' without public representation. Gathering the views of relevant stakeholders, embedding an ethicist with the development team, and resolving conflicts between different specifications of essentially contested concepts create additional work and costs. Unsurprisingly, ethical considerations may be discarded when they conflict with commercial incentives.[22] It cannot be assumed that value-conscious frameworks will be meaningfully implemented in commercial processes that value efficiency, speed, and profit.

### 2.4 Legal and professional accountability

The fourth weakness of a principled approach to AI Ethics is the relative lack of legal and professional accountability mechanisms. Medicine is governed by legal and professional frameworks which uphold professional standards and provide patients with redress for negligent behaviour, including malpractice law, licensing and certification schemes, ethics committees, and professional medical boards.[53,54] Medical institutions subject to regulation help ensure these standards are upheld.[13] Legally supported accountability mechanisms provide an external impetus for health professionals to fulfil their fiduciary duties, amplifies complementary forms of self-governance by establishing a clear link between 'bad' behaviour and professional sanctions (e.g. losing one's license to practice),[11] mandates a professional standard of care, and allows patients to make claims against negligent members of the profession.

Excluding certain types of risks (e.g. privacy violations governed by data protection law), AI development does not have comparable professionally or legally endorsed accountability mechanisms.[62] This is a problem. Serious, long-term commitment to self-regulatory frameworks cannot be taken for granted.[4,5]

Prior research on the impact of codes of ethics on professional behaviour has revealed mixed results. Codes are often followed in letter rather than spirit, or as a 'checklist' rather than as part of a critical reflexive practice.[24,50,63,64] A recent study of the ACM Code of Ethics revealed that it has little effect on the day-to-day decision-making of software engineering professionals and students.[65] Other studies of corporate and professional codes of ethics outside computing have reported similar results.[66–68] A recent meta-analysis of evidence on the impact of codes on professional behaviour found that the mere existence of a code has no discernible effect on unethical behaviour; rather, an effect is only found when codes (and their underlying principles) are embedded in organisational culture and actively enforced.[29,69,70] Norms must be clearly defined and highly visible if they are to influence practitioners[11] and inspire peer self-governance. Current governance structures in AI companies are insufficient in this regard.[2]

External sanctions for breaching a code are also key for adherence and effective self-governance.[11] Compared with medicine, information professions lack sanctions that can impact the professional's livelihood.[24] While software engineering degrees can now be accredited, a license is not required to practice,[71] with some national exceptions.[72–74] Professional bodies such as the IEEE and ACM lack formal sanction powers beyond expulsion from the organisation which, absent licensing, does not impact the ability to practice.

While stronger legal and professional accountability mechanisms could be adopted, this seems unlikely in the near term. AI development is not a unified profession with a long-standing history and harmonised aims. AI developers do not formally provide a public service, meaning public interests need not be given primacy.[75] AI does not operate in a single sector, meaning any new legal or professional mechanisms must account for myriad potential benefits and harms, and integrate with existing sector-specific law. Finally, proposals to introduce professional sanctions and licensing schemes for computing professionals are also not new, but have thus far seen limited uptake.[72,74,76,77]

These weaknesses in existing legal and professional accountability mechanisms for AI raises a difficult question: is it enough to define 'good intentions' and hope for the best? Without complementary punitive mechanisms and governance bodies to 'step in' when self-governance fails, a principled approach runs the risk of merely providing false assurances of ethical or trustworthy AI.[78]

## 3    Where should AI Ethics go from here?

While principlism has undoubtedly significantly influenced medical ethics owing to the four characteristics analysed above, it has not been an unqualified success. The comparative shortcomings of AI development along these lines thus gives cause for concern. First, AI development is not a formal profession with aims that align with public interests. Second, developers are not governed by a historically validated account of what it means to be a 'good' AI developer. Third, outside of academic contexts AI development lacks proven methods to translate principles into practice. And finally, when a developer falls afoul of these vaguely defined requirements, there remain few sanction mechanisms and channels for redress to set things right. Signing up to self-regulatory codes lacking clearly defined and enforceable obligations costs developers nothing, but can have immediate benefits in terms of trustworthiness and reputation. Together, these shortcomings point towards significant challenges facing the implementation of AI Ethics.

We must therefore hesitate to celebrate consensus around high-level principles that hide deep political and normative disagreement. Shared principles are not enough to guarantee trustworthy or ethical AI in the future. Without a fundamental shift in regulation, translating principles into practice will remain a competitive, not cooperative, process. This is a problem, as principles remain vacuous until tested, at which point the true costs and value of a principled approach to AI Ethics will be revealed. Conflicting prescriptions of essentially contested concepts are to be expected. Resolving these conflicts is where the real work starts for AI Ethics. A key question remains: how can this essential work be supported by government, industry, and civil society?

1. **Clearly define sustainable pathways to impact**

A principled approach requires cooperative oversight to ensure translated norms and requirements remain fit for purpose and impactful over time. Going forward, the long-term aims and pathways to impact of principled initiatives must be more clearly defined (see: Table 2 for key questions). Binding and highly visible accountability structures as well as clear implementation and review processes are needed at a sectoral and organisational level.[25] Professional and institutional norms can be established by defining clear requirements for inclusive design, transparent ethical review,[16] documentation of models and datasets, and independent ethical auditing.

2. **Support 'bottom-up' AI Ethics in the private sector**

A 'top-down' approach to AI Ethics is uniquely difficult due to the diversity of technologies described as 'AI'. In such a diverse field, generalist top-down approaches must be complemented by bottom-up case studies of production AI systems. Local practices can be collaboratively assessed to specify principles and define precedents to move professional standards forward.[11,52] Novel cases reveal new challenges for AI Ethics, which are desperately needed to move the field beyond well-worn cases[79,80] and develop sector- and case-specific guidelines, technical solutions, and an empirical knowledge base detailing the impact and harms of production AI technologies. Much current 'bottom-up' work pursues technological solutions and metrics for ethical concepts amenable to quantification (e.g. fairness),[5,81] and occurs primarily in an academic environment. Increased support and access to development settings should be made available to support multi-disciplinary bottom-up research and development in AI Ethics, particularly in commercial development contexts currently closed to external scrutiny.

3. **License developers of high-risk AI**

To encourage long-term recognition of ethical commitments, it may be necessary to formally establish AI development as a profession with equivalent standing to other high-risk professions. It is a regulatory oddity that we license professions providing a public service, but not the profession responsible for developing technical systems to augment or replace human expertise and decision-making within them. The risks of licensed professions have not dissipated, but rather been displaced to AI. To unpack the significant challenges facing licensing such a diverse set of practitioners,[11] initiatives could initially target developers of systems with elevated risk or built for the public sector, such as facial recognition systems designed for policing.

### 4. Shift from professional ethics to organisational ethics

The outputs of many AI Ethics initiatives resemble professional codes of ethics that address design requirements and the behaviours and values of individual professions.[1] The legitimacy of particular applications and their underlying business and organisational interests remain largely unquestioned.[1,82] This approach conveniently steers debate towards the transgressions of unethical individuals, and away from the collective failure of unethical organisations and business models.[27] Developers will always be constrained by the institutions that employ them. To be truly effective, the ethical challenges of AI cannot conceptualised as individual failures. Going forward, AI Ethics must become an ethics of AI businesses and organisations as well.

### 5. Pursue ethics as a process, not technological solutionism

Many initiatives suggest ethical challenges can best be addressed through "technical and design expertise," and address concepts for which technical fixes seem feasible (e.g. privacy, fairness),[1] but rarely propose technical definitions or explanations.[5] Exceptions such as the IEEE's 'Ethically-Aligned Design' initiative do exist, although the impact and uptake of this work in commercial environments remains to be seen.[83] Nonetheless, the rationale seems to be as follows: insufficient consideration of ethics leads to poor design decisions which create systems that harm users.

This attitude is misguided. The promise of AI largely owes to its apparent capacity to replace or augment human expertise. This malleability means AI inevitably becomes entangled in the ethical and political dimensions of vocations and practices in which it is embedded. AI Ethics is effectively a microcosm of the political and ethical challenges faced in society. Framing ethical challenges in terms of design flaws ensures they remain "fundamentally technical, shielded from democratic intervention."[1] It is foolish to assume that very old and complex normative questions can be solved with technical fixes or 'good' design alone. The risk is that complex, difficult ethical debates will be oversimplified to make the concepts at hand computable and implementable in a straightforward but conceptually shallow manner.[80]

Ethics is not meant to be easy or formulaic. Intractable principled disagreements should be expected and welcomed, as they reflect both serious ethical consideration and diversity of thought. They do not represent failure, and do not need to be 'solved'. Ethics is a process, not a destination. The real work of AI Ethics begins now: to translate and implement our lofty principles, and in doing so to begin to understand the real ethical challenges of AI.

### Acknowledgements

The author would like to thank Prof. Sandra Wachter, Prof. Barbara Prainsack, and Prof. Bernd Stahl for their insightful feedback that has immensely improved the quality of this work. Financial support for this work was provided by the Alan Turing Institute (EPSRC) and the British Academy.

## Competing Interests

The author has previously received reimbursement for conference-related travel from funding provided by DeepMind Technologies Limited.

## References

1. Greene, D., Hoffmann, A. L. & Stark, L. Better, Nicer, Clearer, Fairer: A Critical Assessment of the Movement for Ethical Artificial Intelligence and Machine Learning. 10 (2019).

2. Whittaker, M. *et al. AI Now Report 2018*. (AI Now Institute, 2018).

3. Jobin, A., Ienca, M. & Vayena, E. The global landscape of AI ethics guidelines. *Nat. Mach. Intell.* **1**, 389–399 (2019).

4. Nemitz, P. Constitutional democracy and technology in the age of artificial intelligence. *Philos. Trans. R. Soc. Math. Phys. Eng. Sci.* **376**, 20180089 (2018).

5. Hagendorff, T. The Ethics of AI Ethics -- An Evaluation of Guidelines. *ArXiv190303425 Cs Stat* (2019).

6. Calo, R. Artificial Intelligence Policy: A Primer and Roadmap Symposium - Future-Proofing Law: From RDNA to Robots (Part 2). *UC Davis Law Rev.* **51**, 399–436 (2017).

7. Whittlestone, J., Nyrup, R., Alexandrova, A., Dihal, K. & Cave, S. *Ethical and societal implications of algorithms, data, and artificial intelligence: a roadmap for research*. 59 (Nuffield Foundation, 2019).

8. Floridi, L. *et al.* AI4People—An Ethical Framework for a Good AI Society: Opportunities, Risks, Principles, and Recommendations. *Minds Mach.* **28**, 689–707 (2018).

9. OECD. Forty-two countries adopt new OECD Principles on Artificial Intelligence - OECD. *Forty-two countries adopt new OECD Principles on Artificial Intelligence* (2019). Available at: http://www.oecd.org/science/forty-two-countries-adopt-new-oecd-principles-on-artificial-intelligence.htm. (Accessed: 22nd May 2019)

10. High Level Expert Group on Artificial Intelligence. *Ethics Guidelines for Trustworthy AI*. (European Commission, 2019).

11. Filipović, A., Koska, C. & Paganini, C. *Developing a Professional Ethics for Algorithmists: Learning from the Examples of Established Ethics*. (Bertelsmann Stiftung, 2018).

12. Beauchamp, T. L. & Childress, J. F. *Principles of biomedical ethics*. (Oxford University Press, 2009).

13. Bosk, C. L. Bioethics, Raw and Cooked: Extraordinary Conflict and Everyday Practice. *J. Health Soc. Behav.* **51**, S133–S146 (2010).

14. Beauchamp, T. L. & DeGrazia, D. Principles and Principlism. in *Handbook of Bioethics:: Taking Stock of the Field from a Philosophical Perspective* (ed. Khushf, G.) **78**, 55–74 (Springer Science & Business Media, 2006).

15. Marshall, T. H. The recent history of professionalism in relation to social structure and social policy. *Can. J. Econ. Polit. Sci. Can. Econ. Sci. Polit.* **5**, 325–340 (1939).

16. Frankel, M. S. Professional codes: Why, how, and with what impact? *J. Bus. Ethics* **8**, 109–115 (1989).

17. *Black's law dictionary*. (West, 2009).

18. MacIntyre, A. *After Virtue: A Study in Moral Theory*. (Gerald Duckworth & Co Ltd, 2007).

19. Gillon, R. Do doctors owe a special duty of beneficence to their patients? *J. Med. Ethics* **12**, 171–173 (1986).

20. Pellegrino, E. D. & Thomasma, D. C. *The virtues in medical practice*. (Oxford University Press, 1993).

21. Van den Bergh, J., Deschoolmeester, D., Deschoolmeester, D. & Vlerick Leuven Gent Management School, Gent, Belgium. Ethical Decision Making in ICT: Discussing the Impact of an Ethical Code of Conduct. *Commun. IBIMA* 1–11 (2010). doi:10.5171/2010.127497

22. Manders-Huits, N. & Zimmer, M. Values and Pragmatic Action: The Challenges of Introducing Ethical Intel- ligence in Technical Design Communities. **10**, 8 (2009).

23. McDowell, B. *Ethical Conduct and the Professional's Dilemma: Choosing Between Service and Success*. (Quorum Books, 1991).

24. Iacovino, L. Ethical Principles and Information Professionals: Theory, Practice and Education. *Aust. Acad. Res. Libr.* **33**, 57–74 (2002).

25. Boddington, P. *Towards a code of ethics for artificial intelligence research*. (Springer Berlin Heidelberg, 2017).

26. Zarsky, T. Incompatible: The GDPR in the Age of Big Data. *Seton Hall Law Rev.* **47**, (2017).

27. Wachter, S. & Mittelstadt, B. D. A right to reasonable inferences: re-thinking data protection law in the age of Big Data and AI. *Columbia Bus. Law Rev.* **2019**, (2019).

28. Balkin, J. M. Information Fiduciaries and the First Amendment. *UCDL Rev* **49**, 1183 (2015).

29. Kish-Gephart, J. J., Harrison, D. A. & Treviño, L. K. Bad apples, bad cases, and bad barrels: Meta-analytic evidence about sources of unethical decisions at work. *J. Appl. Psychol.* **95**, 1–31 (2010).

30. Wakabayashi, D. & Shane, S. Google Will Not Renew Pentagon Contract That Upset Employees. *The New York Times* (2018).

31. Conger, K. & Wakabayashi, D. Google Employees Protest Secret Work on Censored Search Engine for China. *The New York Times* (2018).

32. Wong, J. C. Demoted and sidelined: Google walkout organizers say company retaliated. *The Guardian* (2019).

33. Parker, D. B. *Ethical Conflicts in Computer Science and Technology*. (AFIPS Press, 1981).

34. Carrese, J. A. & Sugarman, J. The Inescapable Relevance of Bioethics for the Practicing Clinician. *Chest* **130**, 1864–1872 (2006).

35. Brotherton, S., Kao, A. & Crigger, B. J. Professing the Values of Medicine: The Modernized AMA Code of Medical Ethics. *JAMA* **316**, 1041–1042 (2016).

36. Greenfield, B. & Jensen, G. M. Beyond a code of ethics: phenomenological ethics for everyday practice. *Physiother. Res. Int.* n/a-n/a (2010). doi:10.1002/pri.481

37. Panensky, S. A. & Jones, R. Does IT Go Without Saying? *Prof. Times* **Summer 2018**, 30–33 (2018).

38. Perlman, D. T. Who Pays the Price of Computer Software Failure Notes and Comments. *Rutgers Comput. Technol. Law J.* **24**, 383–416 (1998).

39. Mittelstadt, B., Allo, P., Taddeo, M., Wachter, S. & Floridi, L. The ethics of algorithms: Mapping the debate. *Big Data Soc.* **3**, (2016).

40. Jones, T. M. Ethical Decision Making by Individuals in Organizations: An Issue-Contingent Model. *Acad. Manage. Rev.* **16**, 366–395 (1991).

41. Metcalf, J. & Crawford, K. Where are human subjects in Big Data research? The emerging ethics divide. *Big Data Soc.* **3**, 2053951716650211 (2016).

42. Burrell, J. How the Machine 'Thinks:' Understanding Opacity in Machine Learning Algorithms. *Big Data Soc.* (2016). doi:10.1177/2053951715622512

43. Floridi, L. Faultless responsibility: on the nature and allocation of moral responsibility for distributed moral actions. *Philos. Trans. R. Soc. Math. Phys. Eng. Sci.* **374**, 20160112 (2016).

44. Awad, E. *et al.* The Moral Machine experiment. *Nature* **563**, 59 (2018).

45. Ess, C. Ethical pluralism and global information ethics. *Ethics Inf. Technol.* **8**, 215–226 (2006).

46. van den Hoven, J. Computer Ethics and Moral Methodology. *Metaphilosophy* **28**, 234–248 (1997).

47. Gallie, W. B. Essentially Contested Concepts. *Proc. Aristot. Soc.* **56**, 167–198 (1955).

48. Richardson, H. S. Specifying Norms as a Way to Resolve Concrete Ethical Problems. *Philos. Public Aff.* **19**, 279–310 (1990).

49. Turner, L. Bioethics in a Multicultural World: Medicine and Morality in Pluralistic Settings. 19 (2003).

50. Rhodes, R. Good and not so good medical ethics. *J. Med. Ethics* **41**, 71–74 (2015).

51. Clouser, K. D. & Gert, B. A Critique of Principlism. *J. Med. Philos.* **15**, 219–236 (1990).

52. Degrazia, D. Moving Forward in Bioethical Theory: Theories, Caes, and Specified Principlism. **17**, 29 (1992).

53. Orentlicher, D. The Influence of a Professional Organization on Physician Behavior Symposium on the Legal and Ethical Implications of Innovative Medical Technology. *Albany Law Rev.* **57**, 583–606 (1993).

54. Papadakis, M. A. *et al.* Disciplinary Action by Medical Boards and Prior Behavior in Medical School. *N. Engl. J. Med.* **353**, 2673–2682 (2005).

55. Toulmin, S. How medicine saved the life of ethics. *Perspect. Biol. Med.* **25**, 736–750 (1982).

56. van de Poel, I. Translating Values into Design Requirements. in *Philosophy and Engineering: Reflections on Practice, Principles and Process* (eds. Michelfelder, D. P., McCarthy, N. & Goldberg, D. E.) **15**, 253–266 (Springer Netherlands, 2013).

57. Gillon, R. Defending the four principles approach as a good basis for good medical practice and therefore for good medical ethics. *J. Med. Ethics* **41**, 111–116 (2015).

58. Friedman, B., Hendry, D. G. & Borning, A. A Survey of Value Sensitive Design Methods. *Found. Trends® Human–Computer Interact.* **11**, 63–125 (2017).

59. Friedman, B. & Kahn, P. H. Human agency and responsible computing: Implications for computer system design. *J. Syst. Softw.* **17**, 7–14 (1992).

60. Morley, J., Floridi, L., Kinsey, L. & Elhalal, A. From What to How. An Overview of AI Ethics Tools, Methods and Research to Translate Principles into Practices. *ArXiv190506876 Cs* (2019).

61. Shilton, K. Values Levers: Building Ethics into Design. *Sci. Technol. Hum. Values* **38**, 374–397 (2013).

62. MacKinnon, K. S. Computer Malpractice: Are Computer Manufacturers, Service Burreaus, and Programmers Really the Professionals They Claim to Be. *St. Clara Rev* **23**, 1065 (1983).

63. Bynum, T. W. Ethical decision making and case analysis in computer ethics. in *Computer ethics and professional responsibility* (eds. Bynum, T. W. & Rogerson, S.) 60–87 (Blackwell, 2004).

64. Ladd, J. The quest for a code of professional ethics: an intellectual and moral confusion. in *Ethical issues in the use of computers* (eds. Johnson, D. G. & Snapper, J. W.) 8–13 (Wadsworth Publ. Co., 1985).

65. McNamara, A., Smith, J. & Murphy-Hill, E. Does ACM's code of ethics change ethical decision making in software development? in *Proceedings of the 2018 26th ACM Joint Meeting on European Software Engineering Conference and Symposium on the Foundations of Software Engineering - ESEC/FSE 2018* 729–733 (ACM Press, 2018). doi:10.1145/3236024.3264833

66. Brief, A. P., Dukerich, J. M., Brown, P. R. & Brett, J. F. What's wrong with the treadway commission report? Experimental analyses of the effects of personal values and codes of conduct on fraudulent financial reporting. *J. Bus. Ethics* **15**, 183–198 (1996).

67. Helin, S. & Sandström, J. An Inquiry into the Study of Corporate Codes of Ethics. *J. Bus. Ethics* **75**, 253–271 (2007).

68. McCabe, D. L., Trevino, L. K. & Butterfield, K. D. The Influence of Collegiate and Corporate Codes of Conduct on Ethics-Related Behavior in the Workplace. *Bus. Ethics Q.* **6**, 461–476 (1996).

69. Jin, K. G., Drozdenko, R. & Bassett, R. Information Technology Professionals' Perceived Organizational Values and Managerial Ethics: An Empirical Study. *J. Bus. Ethics* **71**, 149–159 (2007).

70. Shilton, K. "That's Not An Architecture Problem!": Techniques and Challenges for Practicing Anticipatory Technology Ethics. 7 (2015).

71. Goertzel, K. M. Legal liability for bad software. *CrossTalk* **23**, (2016).

72. Laplante, P. A. Licensing professional software engineers: seize the opportunity. *Commun ACM* **57**, 38–40 (2014).

73. Pour, G., Griss, M. L. & Lutz, M. The push to make software engineering respectable. *Computer* **33**, 35–43 (2000).

74. Seidman, S. B. The Emergence of Software Engineering Professionalism. in *E-Government Ict Professionalism and Competences Service Science* (eds. Mazzeo, A., Bellini, R. & Motta, G.) **280**, 59–67 (Springer US, 2008).

75. Abbott, A. Professional Ethics. *Am. J. Sociol.* 32 (1983).

76. O'Connor, J. E. Computer Professionals: The Need for State Licensing. *Jurimetr. J.* **18**, 256–267 (1978).

77. Gotterbarn, D. The Ethical Computer Practitioner—Licensing the Moral Community: A Proactive Approach. *SIGCSE Bull* **30**, 8–10 (1998).

78. Suchman, L. Corporate Accountability. *Robot Futures* (2018).

79. Angwin, J., Larson, J. & Kirchner, L. Machine Bias. *ProPublica* (2016).

80. European Group on Ethics in Science and New Technologies. *Statement on Artificial Intelligence, Robotics and 'Autonomous' Systems*. (European Commission, 2018).

81. Holstein, K., Vaughan, J. W., Daumé III, H., Dudík, M. & Wallach, H. Improving fairness in machine learning systems: What do industry practitioners need? *ArXiv181205239 Cs* (2018). doi:10.1145/3290605.3300830

82. Benkler, Y. Don't let industry write the rules for AI. *Nature* **569**, 161–161 (2019).

83. The IEEE Global Initiative on Ethics of Autonomous and Intelligent Systems. *Ethically Aligned Design*. (IEEE, 2019). doi:10.1007/978-3-030-12524-0_2

**Tables**

**Table 1 - Characteristics of a Formal Profession**

1. **Specialised education and training -** Members are expected to have undertaken extensive specialised education and training,[51] typically in accredited degree programmes.[33]
2. **Commitment to public service** - Professions involve a public declaration to provide a service to society or for the public good making use of specialised, often privileged expertise[65] which takes precedence over individual gain.[51]
3. **Higher standard of care -** Professionals commit to upholding higher ethical standards than would normally be expected in business relationships in service to both the client and the public.[51]
4. **Enforcement and self-governance -** Often, these standards are recorded in an ethical code and enforced through a disciplinary system, administered by professional associations.[63,65]
5. **Licensing** – Entry to the profession is restricted by (government sanctioned) licensure to highly skilled individuals as a means to protect the public.[33,51]

| Table 2 - Key questions to assess AI Ethics value and principle statements |
|---|
| 1. Who wrote it, and how? |
| 2. Who is it intended for, and what is its purpose? |
| 3. Why should I follow it? |
| 4. How do I follow or implement it? |
| 5. How should I resolve conflicting interpretations of essentially contested concepts? |
| 6. How will you know I am following it? |
| 7. What happens if I fail to follow it? |
| 8. How can I raise disagreements or questions for clarification? |